# *N*-1 Reduced Optimal Power Flow Using Augmented Hierarchical Graph Neural Network

Thuan Pham, *Student Member, IEEE* and Xingpeng Li, *Senior Member, IEEE*

*Abstract*— **Optimal power flow (OPF) is used to perform generation redispatch in power system real-time operations. *N*-1 OPF can ensure safe grid operations under diverse contingency scenarios. For large and intricate power networks with numerous variables and constraints, achieving an optimal solution for real-time *N*-1 OPF necessitates substantial computational resources. To mitigate this challenge, machine learning (ML) is introduced as an additional tool for predicting congested or heavily loaded lines dynamically. In this paper, an advanced ML model known as the augmented hierarchical graph neural network (AHGNN) was proposed to predict critical congested lines and create *N*-1 reduced OPF (*N*-1 ROPF). The proposed AHGNN-enabled *N*-1 ROPF can result in a remarkable reduction in computing time while retaining the solution quality. Several variations of GNN-based ML models are also implemented as benchmark to demonstrate effectiveness of the proposed AHGNN approach. Case studies prove the proposed AHGNN and the associated *N*-1 ROPF are highly effective in reducing computation time while preserving solution quality, highlighting the promising potential of ML, particularly GNN in enhancing power system operations.**

## NOMENCLATURE

| | |
|---|---|
| $G$ | Set of online generators. |
| $K$ | Set of lines. |
| $N$ | Set of buses. |
| $C$ | Set of contingency cases. |
| $c_g$ | Cost of generation for generator unit $g$. |
| $P_g$ | Generation dispatch of generator unit $g$. |
| $P_g^{res}$ | Reserve generation of generator unit $g$. |
| $P_g^{min}$ | Minimum generation of generator unit $g$. |
| $P_g^{max}$ | Maximum generation of generator unit $g$. |
| $P_k$ | Power flow of line $k$. |
| $x_k$ | Reactance of line $k$. |
| $\theta_{f(k)}$ | Angle of bus from line $k$. |
| $\theta_{t(k)}$ | Angle of bus to line $k$. |
| $RateA_k$ | Line rating limit of line $k$ in the base. |
| $RateC_k$ | Line rating limit of line $k$ in the contingency case. |
| $d_n$ | Load profile at bus $n$. |
| $N1_{ck}$ | Status of each line $k$ per contingency case (binary). |
| $x_i$ | Node's features of node $i$ in a graph. |
| $e_{(ij)}$ | Edge's features between node $i$ and node $j$ in a graph. |
| $s_{ij}$ | Stack of features from both node and edge. |
| $y_i$ | Truth label of prediction. |
| $\hat{y}_i$ | Probabilities of the prediction. |

Thuan Pham and Xingpeng Li are with the Department of Electrical and Computer Engineering, University of Houston, Houston, TX, 77204, USA. (e-mail: tdpham7@cougarnet.uh.edu; xli83@central.uh.edu).

## I. INTRODUCTION

Optimal power flow (OPF) constitutes a prevalent technique for power system operations, analysis, and scheduling. It encompasses the optimization of an objective function that can assume various forms while complying with a set of operational and physical constraints. Security constraints must be enforced in OPF to address power system uncertainties, of which line contingency due to unforeseen circumstances is quite common in practice and may cause very severe outcomes. The applications of OPF have expanded due to its capacity to cope with diverse aspects of the power system network. *N*-1 OPF is used to address sudden major disturbances, particularly line outages, and ensure the power grid is able to operate safely and serve all the demand in each possible post-contingency scenario [1] [2]. With *N*-1 OPF, power system reliability can be enhanced as compared to OPF with only reserve-based reliability requirement constraints.

Machine learning (ML) constitutes a field within artificial intelligence and computer science dedicated to harnessing data and algorithms to simulate human learning processes, with the goal of progressively enhancing its accuracy over time. The adoption of ML technologies in the power system domain has been growing very fast in recent years. For example, long-term recurrent convolutional network (LRCN) models have been developed for electricity price prediction [3] [4], load forecasting [5], and synchronous inertia estimation [6]. Deep neural network (DNN) and graph neural networks (GNN) have been proposed to predict power flow results, offering an alternative to traditional methods [7]. Logistic regression and multiple deep learning methods have been employed to determine the commitment status of a subset of generators with confidence, which can accelerate the computing process of solving day-ahead scheduling models that select the most cost-effective generation units [8] [9].

There are some prior efforts that study the use of ML for OPF in the literature. In [10], a DNN model was used to predict generations from load inputs in a security-constrained OPF model. In [11], a convolutional neural network (CNN) model was built to predict generation dispatch given load profiles. However, both methods do not account for the topology of the power network. There is inherent information related to the global context of the network that can be used to train and enhance the ML model. Network topology is crucial and correlates to how well a learning model performs under different conditions [12].

GNN that was designed to address network-structured data can utilize network topology as part of the features during the training process [13]. GNN is an advanced neural network model that captures the dependence of graphs via message passing between the nodes of graphs. The GNN model relies heavily on network topology to pass information from nearby nodes and edges across multiple GNN layers. In [14], it presents a comprehensive overview of GNN for power system applications such as fault detection, time-series prediction, and power flow calculation. Different variants of GNN have been evaluated in [15] for possible applications in power system control and optimization. As an example, one variant of GNN has been used as a prediction tool for the power flow model [16]. Recently, another variant of GNN has been used to predict wind speed for generation at wind farm clusters [17]. GNN has the potential to expand and improve upon existing tools for power system optimization such as OPF.

In both [18] and [19], network topology is a key factor in calculating OPF using ML. It shows that utilizing the topology as an additional feature will enhance prediction for OPF calculation. A recent attempt to use GNN for OPF calculation did not account for power flow and ignored the line rating limit in its OPF formulation [20]. There were attempts to compute AC OPF using an unsupervised GNN model or deep reinforcement learning [21] [22]; however, constraint errors were still present in their solutions.

In our previous work [23], we proposed an ML algorithm using GNN to assist efficient OPF formulation. Using GNN, we created a reduced OPF (ROPF) model that decreases computing time by 16% compared to full OPF, while the ROPF solutions satisfy all the constraints.

In this paper, we proposed to use GNN for the *N*-1 OPF problem, which evaluates possible *N*-1 contingency scenarios along with the base scenario where no contingency occurs. For the *N*-1 OPF model, the numbers of constraints and variables increase exponentially and are significantly higher compared to the base-case OPF, which requires advanced methods to efficiently solve *N*-1 OPF. Particle swarm optimization [24] and compensation method [25] are two notable methods to simplify the *N*-1 OPF model. However, the particle swarm optimization method did not consider computing time as one of its evaluation metrics while the compensation method did not consider all possible contingencies or objective total cost in its result.

To implement the *N*-1 OPF algorithm using predictions from ML, a sophisticated and highly complex GNN model is needed to predict congested lines for the base case and the *N*-1 contingency cases. In this paper, we proposed an augmented hierarchical graph neural network (AHGNN) to effectively identify and eliminate unnecessary non-binding constraints for *N*-1 OPF. Previous applications of hierarchical models in deep learning, such as classifications of bioimaging data, produced significant improvement in accuracy compared to a straightforward flat ML model [26] [27]. A hierarchical deep learning model was also recently used for battery degradation predictions, which demonstrated better inference performance than single-stage ML models [28].

The proposed AHGNN model is a nested two-stage hierarchical ML model. The first-tiered GNN model, base-case GNN (BC-GNN), is trained to predict congested lines in the normal pre-contingency base case. These predictions from this first BC-GNN model serve a dual purpose: (i) identify a subset of critical lines for further monitoring and (ii) are utilized as crucial features for the second-tiered GNN model that is a contingency-case GNN (CC-GNN) model. This CC-GNN model is employed to predict congested lines for each possible *N*-1 contingency case. These predictions collectively aim to reduce the number of monitoring lines or constraints related to line flow limits in the *N*-1 contingency cases. Using these predictions, the full *N*-1 OPF problem is effectively transformed into an *N*-1 reduced OPF, or *N*-1 ROPF problem. Using the trained GNN models on samples on the validation data set, predictions were made on the base cases and the *N*-1 contingency cases. Subsequently, those predictions were used to create *N*-1 ROPF. The solutions to the *N*-1 ROPF were then checked for all line thermal limit constraints for both base-case and contingency-cases, identifying a set of the notable congested lines (NCL) that is then used as another source to identify critical line for solving future *N*-1 ROPF problems.

To assess the effectiveness of the proposed AHGNN's predictive capabilities, we also established four additional ML models, using GNN as the main architecture, for comparative analysis, examining the validity of their predictions and performance against the full *N*-1 OPF. These five ML models were created using a combination of various predictions for the base cases, contingency cases, and notable congested lines. The ideal model should have predictions that result in zero constraint violations for the *N*-1 ROPF solutions while the objective total costs should be nearly identical to the full *N*-1 OPF solutions. With the proposed AHGNN model, a significant reduction in computational time required to find optimal solutions is achieved.

Furthermore, we delve into enhancements in hyperparameter tuning, especially loss attenuation during the training phase of the GNN model. We implemented a weighted factor (WF) to the loss function to limit predictions that are more likely to lead to violations of constraints. We perform sensitivity analysis to select the best weighted factor for the proposed GNN model. Predictions using the new loss function showed a remarkable decrease in the violation of constraints for the *N*-1 ROPF solutions.

The remaining sections of this paper are organized as follows. Section II provides a comprehensive introduction to *N*-1 OPF. Section III introduces GNN and explores the architectural layers. Section IV explains the methodologies employed in our research. Section V presents the simulation results and analysis. In Section VI, we draw conclusions based on our findings, summarize the key takeaways, and discuss the implications of our research.

## II. *N*-1 OPTIMAL POWER FLOW

The OPF model is commonly used in practical power systems for real-time economic dispatch. It optimizes the generation dispatch points while satisfying multiple constraints throughout the transmission network and focuses on minimizing the total operating cost. For the OPF model presented below, the objective function is to minimize the total system generation cost as shown in (1). Constraints (2) to (4) describe generation limits as well as generation reserve requirement that is set at 5% of the total load. Line flows are calculated in (5) and are subject to line thermal capacity limits in (6). The nodal power balance is enforced in (7).

$$min \sum_n c_g P_g \qquad\qquad g \in G \qquad (1)$$

$$\sum_n P_g^{res} = 0.05 \times \sum_n d_n \qquad g \in G, n \in N \quad (2)$$

$$P_g^{min} \leq P_g \leq P_g^{max} - P_g^{min} \qquad g \in G \qquad (3)$$

$$P_g^{res} \geq 0 \qquad\qquad g \in G \qquad (4)$$

$$P_k = \frac{\theta_{f(k)} - \theta_{t(k)}}{x_k} \qquad\qquad g \in G \qquad (5)$$

$$-RateA_k \leq P_k \leq RateA_k \qquad k \in K \qquad (6)$$

$$\sum_n P_g + \sum_{n(f)} P_k + \sum_{n(t)} P_k = d_n \qquad n \in N \quad (7)$$

To comply with the *N*-1 reliability requirement, the impacts of various single-element contingencies are evaluated and considered in OPF, creating the *N*-1 OPF model. To be consistent with the industry practices [29] [30], radial line contingencies that will lead to network islanding are not considered in the *N*-1 OPF model in this work. Constraints (8)-(9) ensure the updated generators' power outputs under contingency are still within the associated limits and do not violate the ramping rate limits. Post-contingency line flow equations and thermal limit constraints for each *N*-1 contingency case are shown in (10)-(11) respectively. Equation (12) ensures the post-contingency nodal power balance for each *N*-1 contingency case.

$$P_g^{min} \leq P_{cg} \leq P_g^{max} \qquad c \in C, g \in G \quad (8)$$

$$-P_g^{ramp} \leq P_{cg} - P_g \leq P_g^{ramp} \qquad c \in C, g \in G \quad (9)$$

$$P_{ck} = \frac{\theta_{f(ck)} - \theta_{t(ck)}}{x_k} \times N1_{ck} \qquad c \in C, k \in K \quad (10)$$

$$-RateC_k \leq P_{ck} \leq RateC_k \qquad c \in C, k \in K \quad (11)$$

$$\sum_n P_{cg} + \sum_{n(f)} P_{ck} + \sum_{n(t)} P_{ck} = d_n \quad c \in C, k \in K \quad (12)$$

For practical independent system operators (ISOs), the line thermal limit constraints will be applied only on a small subset of lines that are heavily loaded or congested. These subsets of lines are chosen based on historical load profiles and line loading levels. By keeping an eye on these congested lines and zeroing in on the subsets of heavily loaded or congested lines, ISOs can ascertain that the power system network is operating efficiently while ensuring grid reliability. For *N*-1 OPF problems, the optimal solutions may be constrained by congested lines in both the base case and the *N*-1 cases. These congested lines are considered critical since they restrict the solution within a bounded feasible region. Uncongested lines are not monitored since the optimal solution is not affected by these variables.

$$-RateA_r \leq P_r \leq RateA_r \qquad c \in C, r \in R \quad (13)$$

$$-RateC_r \leq P_{cr} \leq RateC_r \qquad c \in C, r \in R \quad (14)$$

By replacing (6) and (11) with (13) and (14) respectively, the updated *N*-1 OPF model can monitor only a small number of critical congested lines and ignore other non-critical uncongested lines in both the base case and contingency cases. As a result, the majority of line rating constraints could be eliminated from the OPF model, which will substantially reduce the model size and complexity.

### III. Graph Neural Network

Machine learning is a well-established computer algorithm that has been extensively researched and applied in power system optimization. ML models have the remarkable ability to automatically improve and learn from vast amounts of data. These models are trained using data samples to make predictions or decisions without explicit programming. Among popular ML models are CNN and DNN, which have found broad applications across various disciplines. However, these models often overlook the importance of considering the network topology in their predictions.

For electrical networks, the topology contains crucial but implicit information and features that could significantly enhance the learning process. This is where the GNN model holds a distinct advantage over both CNN and DNN models. The GNN model utilizes an adjacency matrix (a matrix of size $nb$ by $nb$, where $nb$ represents the number of buses) to represent the network topology. Furthermore, the GNN model leverages the global context of the network to establish relationships between multiple node-level and edge-level features. In essence, the GNN model allows an optimizable transformation on all attributes of the graph, preserving graph symmetries, and permutation invariance.

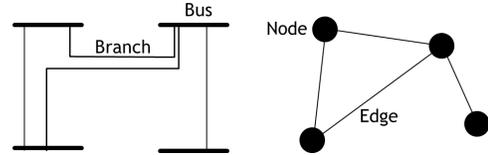

Fig. 1. Transformation of bus-branch model into a graph format.

Fig. 1 above illustrates the transformation of a simple power system's bus-branch model into a graph format. Each bus in the network is represented as a node, holding information such as generation and load profiles as node features. Line flow and line thermal ratings are represented as edge features. The adjacency matrix effectively characterizes the connections between each bus and its neighboring buses. GNN has been used in diverse applications, serving various purposes in node-level, edge-level, and graph-level tasks, depicted in Fig. 2 below [31]-[33].

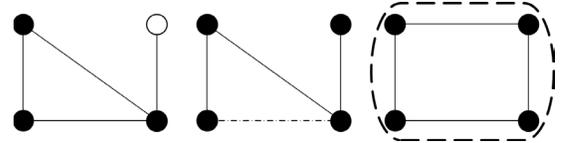

Fig. 2. Node-level vs. Edge-level vs. Graph-level classification task.

During the training stage of the GNN model, (15) below describes the relationship between each node and its neighboring nodes in each forward pass of the training stage. $h_i^k$ contains all the feature information of the current node, while $W_{self}^k$ and $W_{neigh}^k$ are node-wise shared weight matrices. $AGG\left(h_j^{k-1}\right)$ is an aggregation function that combines feature information over all neighbor nodes, $\Omega_i$. For each forward pass, the information acquired from each neighboring node is aggregated and passed through to the next training step. With a well-built GNN model with enough layers, feature information across the entire network can be utilized to enrich the prediction capability of the model. By leveraging the topology of the network during the training process, we can uncover valuable insights and improve the efficiency of power system optimization through GNN.

$$h_i^k = \sigma\left(W_{self}^k \cdot h_i^{k-1} + W_{neigh}^k \cdot AGG\left(h_j^{k-1}, \forall j \in \Omega_i\right)\right) \quad (15)$$

For this work, the XENet layer was utilized as a graph convolutional layer and an integral part of the proposed GNN model. XENet layer was chosen due to its ability to convolve over both edge and node tensors [34]. The XENet layer is a message-passing layer that simultaneously accounts for both the incoming and outgoing information of neighboring nodes, such that a node's representation is based on the messages it receives as well as those it sends. The architecture of the XENet is described using the equations below.

$$s_{ij} = \varphi^{(s)}(x_i||x_j||e_{(i,j)}||e_{(j,i)}) \quad (16)$$

$$s_i^{out} = \sum_{j \in N(i)} a^{out}(s_{ij}) \cdot s_{ij} \quad (17)$$

$$s_i^{in} = \sum_{j \in N(i)} a^{in}(s_{ij}) \cdot s_{ji} \quad (18)$$

$$x'_i = \varphi^{(n)}(x_i||s_i^{out}||s_i^{in}) \quad (19)$$

$$e'_{(i,j)} = \varphi^{(e)}(s_{ij}) \quad (20)$$

where $\varphi^{(s)}$, $\varphi^{(n)}$, $\varphi^{(e)}$ are multi-layer perceptrons with parametric rectified linear unit as the nonlinear activations, and $a^{out}$ and $a^{in}$ are two dense layers with sigmoid activations and a single scalar output.

For a given graph, vector $x_i$ contains the features associated with node $i$, and vector $e_{(i,j)}$ contains the features belonging to edge between node $i$ and node $j$. The XENet layer aggregates information of the feature stacks $s_{ij}$ in (16)-(18) for message-passing between each forward pass. By concatenating the node and edge attributes associated with the incoming and outgoing messages of each node in (16), the multi-layer perceptron $\varphi^{(s)}$ learns to process information such that a node's representation is based on the messages it receives as well as those it sends. The feature stacks are used to aggregate incoming/outgoing information, using self-attention to compute a weighted sum, in (17) and (18) respectively. The incoming/outcoming messages are concatenated and used to update the node attributes, $x'_i$, of the graph in (19). Finally, some additional processing of the feature stacks through $\varphi^{(e)}$ in (20) allows us to compute new edge attributes, $e'_{(i,j)}$, that are dependent on the message exchange between nodes.

XENet is a critical part of our GNN model's architecture. XENet allows us to pay special attention to both incoming and outgoing edge attributes. It enables us to perform edge-levels predictions across the network by stacking features of nodes and edges, and then passing those learned features through each forward pass. The proposed AHGNN method relies on the XENet layer to predict congested lines accurately in order to remove unnecessary constraints. It enables us to create a computationally efficient N-1 ROPF model that decreases the solving time to find the optimal solution.

## IV. METHODOLOGIES

This section explains the proposed AHGNN method that can dynamically determine the monitor lines for the base case and contingency cases: the offline training stage and online prediction stage are presented in two subsections separately. A detailed explanation of the data generation process is also provided. In addition, four benchmark methods are also defined to gauge the proposed AHGNN-based critical line selection approach.

### A. Data Generation:

A large number of samples were generated to represent the historical grid data, which were used to train the proposed ML models. We ran N-1 OPF simulations on the test power system with different load profiles to collect enough samples, which is 10,000 in this work. The load profile of each sample is varied using two random variable parameters. The first parameter shifts the entire load profile up or down within a range of ±5% of the initial load profile. The second parameter shifts load profile at each individual bus within a range of ±2%. The combination of both parameters provides varying load profile curves.

Once the samples had been generated, we solved full N-1 OPF and obtained the solution for each sample. From the solutions, we labeled line flows from all branches based on a pre-defined threshold: the lines whose line flows are above this threshold are considered critical, which will be included in the line monitor set for N-1 ROPF.

In the IEEE 73-bus system, there are a total of 108 branches, of which 2 branches are radial lines. Radial lines are branches that, once removed, will cause network separation. As illustrated in Fig. 3, the two red lines connecting nodes 30 and 54 (blue dot) are radial lines. In the full N-1 OPF model, all non-radial line contingency cases, in addition to the base case, are considered. Thus, the base case and 106 contingency cases are simulated and modeled in N-1 OPF problem. Using ML, congested lines will be predicted for each case of the N-1 ROPF problem.

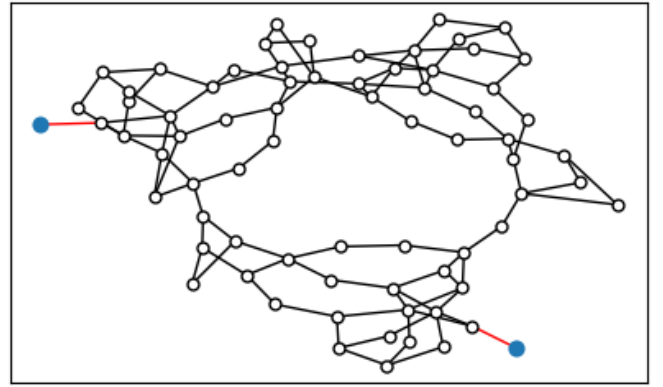

Fig. 3. Graph layout of IEEE 73-bus system.

From the initial 10,000 collected data samples, for the OFFLINE – Training mode, we separated it into three data set:
- Training data set for BC-GNN model
- Validation data set for BC-GNN model
- Testing data set for BC-GNN model.

The testing data set for BC-GNN is further split in two for:
- Training data set for CC-GNN model
- Validation data set for CC-GNN model

The following sections will go into further detail on how each data set is used to train separate GNN models for predictions of congested lines in base cases and contingency cases. Additional 1,000 new samples are generated and used for the ONLINE - Prediction mode to evaluate the proposed AHGNN model as explained below.

## B. OFFLINE – Training Mode:

Fig. 4 illustrates the flowchart for training the two proposed ML models: (i) BC-GNN model for classification of congested lines in the base cases and (ii) CC-GNN model for classification of congested lines in the contingency cases. Once all the data samples are generated, processed, and labeled, they will be divided into three separate data sets for training, validation, and testing. During the OFFLINE training stage, these three data sets will be used to train and evaluate the BC-GNN model to identify critical lines to be monitored for the base case of OPF, where no contingency is considered as part of the input features.

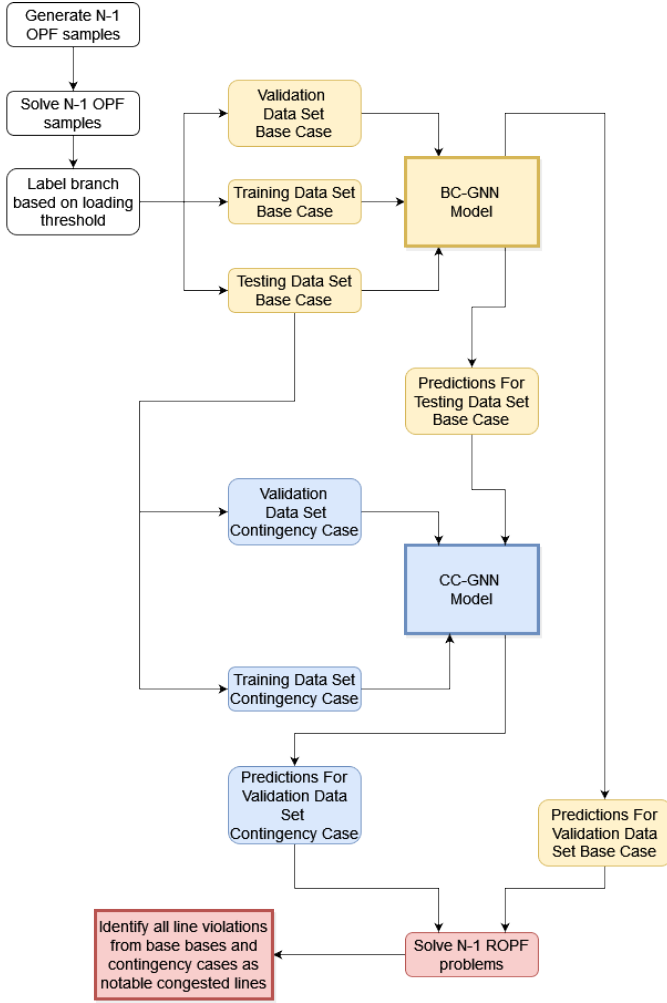

Fig. 4. Flowchart for the OFFLINE – Training mode.

Once the BC-GNN model is trained, prediction results from the testing data set are used as features for the CC-GNN model of the *N*-1 contingency cases. Since each sample contains data for both the base case and the contingency cases, samples from the testing data set of the base cases are split into training and validation data sets of the contingency cases to train the CC-GNN model. By using output from BC-GNN as features for CC-GNN, we were able to enhance the prediction result of the second model CC-GNN. In essence, it is a nested two-stage training process. Finally, predictions are made on the *N*-1 cases validation data set using the trained CC-GNN model. These CC-GNN predictions combined with the BC-GNN predictions made on the same data set are then used to solve

*N*-1 ROPF problems. Congested lines from the solutions of *N*-1 ROPF problems are then identified as notable congested lines for the ONLINE prediction mode.

## C. ONLINE – Prediction Mode:

Once the models are trained, they can be deployed for online prediction. A separate data set of 1,000 new samples that were never used during the training stage is used here to evaluate the effectiveness of both BC-GNN and CC-GNN models. The outputs generated by the BC-GNN model are incorporated as part of the features for the CC-GNN model. Furthermore, NCLs identified from past *N*-1 ROPF solutions during the OFFLINE training stage will also be included in *N*-1 ROPF (online) as monitor lines for new test samples.

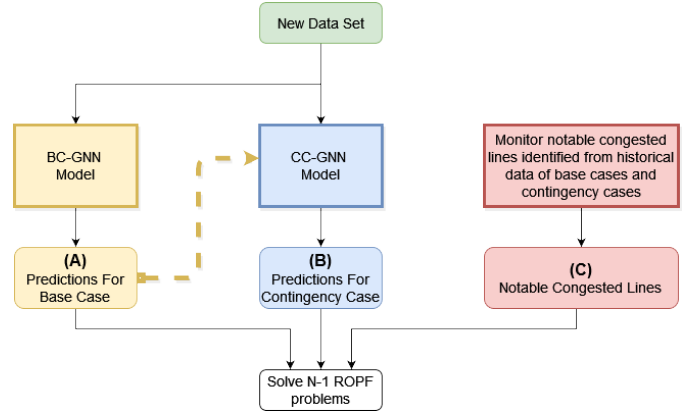

Fig. 5. Flowchart for the ONLINE - Prediction mode.

Fig. 5 summarizes how three distinct processes to identify critical lines, two use ML models and one uses previous solutions of *N*-1 ROPF, to only include possibly congested lines in the *N*-1 OPF problem. This strategic culling of uncongested lines plays a pivotal role in substantially reducing the number of constraints, consequently culminating in the enhancement of the *N*-1 ROPF model's efficacy. To gauge the effectiveness of our proposed approach, we establish a comparative benchmark by developing four other models for critical monitor line selection:

- Notable Congested Lines [NCL]: only use the notable congested lines identified offline [(C) as shown in Fig. 5].
- Graph Neural Network [GNN]: uses predictions from base case and *N*-1 cases, but without BC-GNN predictions as additional features for CC-GNN [(A) and (B), as shown in Fig. 5].
- Augmented Graph Neural Network [AGNN]: uses predictions from base case and *N*-1 cases [(A) and (B)], but without BC-GNN predictions as additional features for CC-GNN, along with notable congested lines [(C)].
- Hierarchical Graph Neural Network [HGNN]: uses predictions from base case and *N*-1 cases only, and with BC-GNN predictions as additional features for CC-GNN [(A) and (B)].
- Augmented Hierarchical Graph Neural Network [AHGNN] – the proposed method: uses predictions from base case and *N*-1 cases, and with BC-GNN predictions as additional features for CC-GNN, along with notable congested lines [(A), (B), (C)].

### D. Loss Factor Function:

Using an advanced GNN model for the prediction of line congestions, the incorrected predictions will result in two possible types of errors: (i) type I false positive error – incorrectly predicting a lightly loaded line to be heavily loaded; and (ii) type II false negative error – incorrectly predicting a heavily loaded line to be lightly loaded. Having more type II errors will decrease the size of the subset of critical lines that we need to watch for $N$-1 ROPF, which in turn will likely speed up the solving time. However, incorrect predictions leading to type II errors have a direct correlation to the number of constraint violations in the final optimal solution. Thus, we want to enhance the GNN model to lower the number of type II errors, which in turn will dramatically reduce the number of constraint violations.

$$Loss = -\sum_{i=1}^{N} y_i \times \log(\hat{y}_i) \qquad (19)$$

$$WF = [1, \ldots, R] \qquad R > 0 \qquad (20)$$

$$WF\ Loss = -\sum_{i=1}^{N} y_i \times \log(\hat{y}_i \times WF) \qquad (21)$$

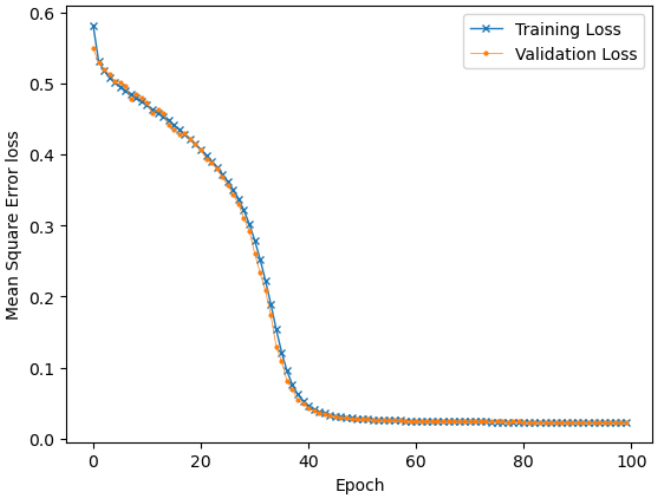

Fig. 6. MSE loss for training data set vs. validation data set during the training phase of the BC-GNN model.

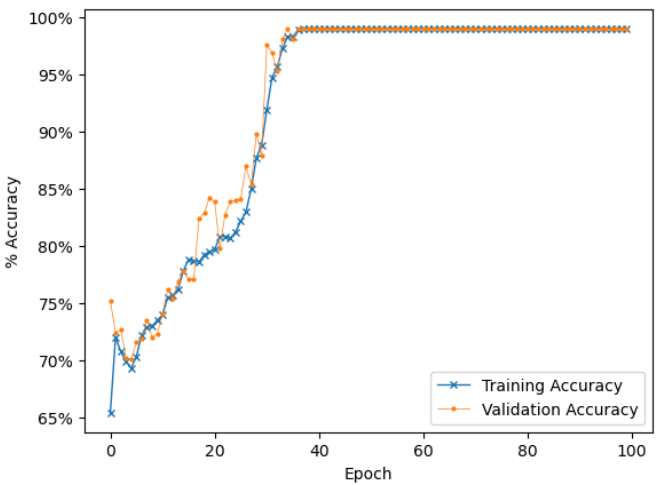

Fig. 7. Percent accuracy of prediction for training data set vs. validation data set during the training phase of the BC-GNN model.

To improve the accuracy level of prediction, we need to significantly decrease the loss level during the training process. (19) describe the typical categorically cross-entropy loss function, where $y_i$ is the truth label and $\hat{y}_i$ is the probabilities of the line binding status. We introduce the term weighted factor, $WF$, in (20). It is a relative ratio between 0 and 1 for each class in the loss function. By multiplying the term $WF$ in (21), we created a new loss equation that emphasizes the loss value of one class over another. Essentially, the loss value will be propagated throughout the model and minimize the prediction error of type II over type I. By having fewer type II errors and more type I errors, the number of constraints in the $N$-1 ROPF model will increase, but the probability of having line constraint violations that lead to infeasible OPF solutions will significantly decrease.

## V. RESULTS AND ANALYSIS

GNN models and $N$-1 ROPF algorithm were developed in Python. We used the Pyomo library [35] [36] and Gurobi solver to solve $N$-1 ROPF problems. The GNN models were built using the spektral library [37]. All figures and graphs were generated using matplotlib [38] and networkx [39]. A computer with 4 cores i7-4790K CPU, GTX3090 graphic card, and 32GB memory was utilized for all coding activities.

### A. Model Training:

From Fig. 6 and 7, the BC-GNN model performed well during the training phase. The curve for training loss tracked well with validation loss. The loss plot has a smooth decline that leads to a flat plateau. Percent accuracy of prediction has a similar, but inverse appearance with a nearly 99% accuracy for the base case model. Both figures suggested that the training process went well where maximum accuracy is achieved while minimizing loss. The curves for loss plot and accuracy plot of CC-GNN model exhibit the same appearance as BC-GNN and are not shown here.

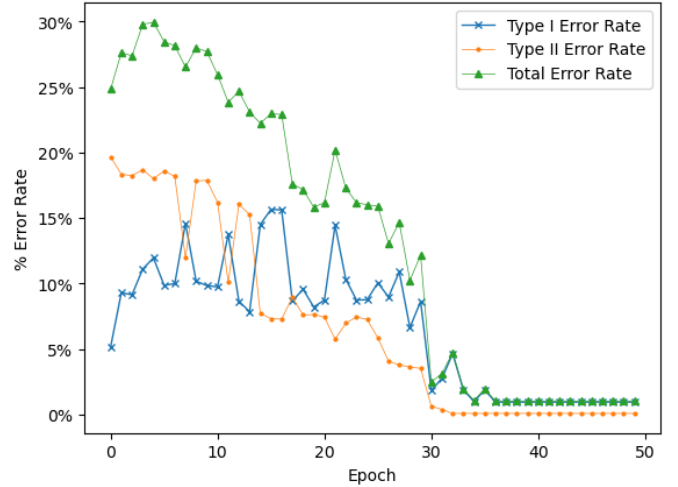

Fig. 8. Percent of error rate during the training phase of the BC-GNN model.

### B. Loss Factor function:

By implementing the WF loss function, a significant decrease in type II errors within our baseline prediction was observed. As depicted in Fig. 8, the incidence of type II errors dropped from 0.358% to 0.059%, constituting a remarkable six-fold reduction in magnitude. Nevertheless, the overall percentage of incorrect predictions, encompassing both type I and type II errors, remained relatively unchanged. Fig. 9 illustrates our tracking of both error types across the first 50 epochs of the training process. Notably, while the type II error rate exhibited a high initial value, it substantially decreased in the initial 30 epochs and subsequently stabilized by the 50th epoch. This observation underscores the efficacy of our new loss function in adeptly propagating gradient loss during training, effectively curtailing the error rate pertaining to one specific class. Our hypothesis posits that the emphasis on the loss value of one class over the other facilitated a reduction in prediction errors for that class while keeping the total error rate at the same level. The outcomes of our analysis affirm our success in markedly mitigating type II error predictions, consequently leading to a lower frequency of line rating violations within the $N$-1 ROPF solutions.

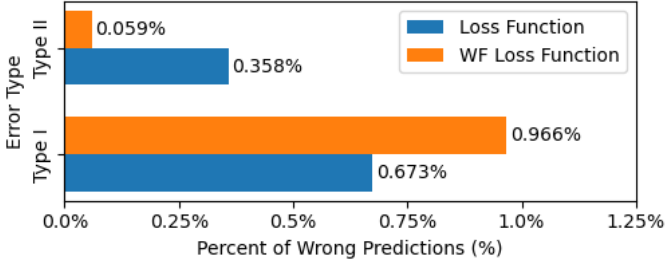

Fig. 9. Comparison of percentage error between type I and type II using the typical loss function vs. the (Weighed Factor) WF loss function.

### C. The Proposed AHGNN Model:

Upon obtaining predictions for line congestion, these forecasts are integrated into the $N$-1 ROPF problem. Once solutions have been found for the $N$-1 ROPF problem, a thorough validation process was used to ensure that all stipulated constraints have been met. Remarkably, our proposed model, AHGNN, outperformed all the other benchmark models by a significant margin. This superior performance is illustrated in Fig. 10, where we present a stacked bar chart that contrasts the incidence of line rating violations per sample within the $N$-1 post-contingency scenarios, utilizing a data set of 1000 samples. Among the five models examined, it is noteworthy that only the AHGNN model boasts zero line rating violation across all solution instances for each sample.

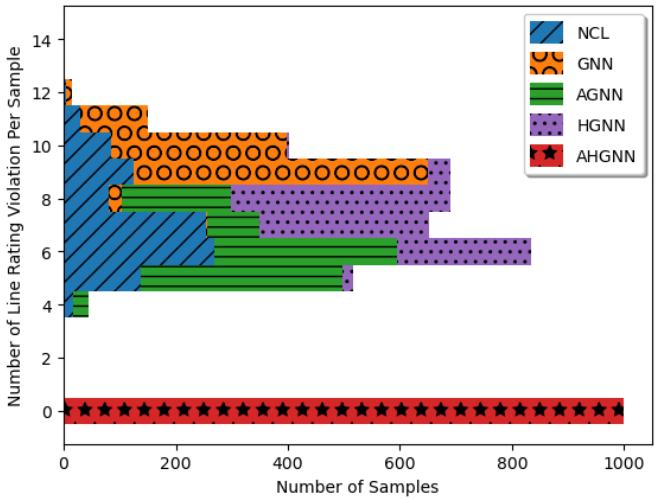

Fig. 10. Comparison of the number of line rating violations for the five proposed models.

The four alternative benchmark models encountered challenges in effectively solving $N$-1 ROPF solutions without violations. A detailed breakdown of the frequency of line rating violations for each model is presented in Table I. Among the other four benchmark models, instances of $N$-1 ROPF solutions consistently violated line rating constraints were observed, ranging from a minimum of 4 lines to a substantial number of 14 lines. The tabulated data strongly suggests that employing isolated prediction sources, without integration, fails to yield solutions of sufficient quality that adhere to all modeling constraints. Notably, within the GNN model, reliance solely on predictions from the base and $N$-1 cases is associated with the highest count of line rating violations compared to other models. This is particularly evident in over 500 samples that experience 9 or more line rating violations. Despite the incorporation of base case predictions to enhance $N$-1 predictions in the HGNN model, the solutions still fall short of desired expectations.

Conversely, the AHGNN model, which incorporates NCL into its predictive process, stands as the sole exception. This model consistently generates solutions free from any constraints or line rating violations, underscoring its effectiveness and superiority in producing robust and accurate $N$-1 ROPF solutions.

TABLE I
Line rating violations for the five proposed models per 1000 samples.

| | NCL | GNN | AGNN | HGNN | AHGNN |
|---|---|---|---|---|---|
| 0 | 0 | 0 | 0 | 0 | 1000 |
| 1 | 0 | 0 | 0 | 0 | 0 |
| 2 | 0 | 0 | 0 | 0 | 0 |
| 3 | 0 | 0 | 0 | 0 | 0 |
| 4 | 16 | 0 | 28 | 1 | 0 |
| 5 | 138 | 0 | 359 | 19 | 0 |
| 6 | 269 | 0 | 326 | 239 | 0 |
| 7 | 254 | 3 | 92 | 304 | 0 |
| 8 | 81 | 23 | 195 | 391 | 0 |
| 9 | 124 | 525 | 0 | 42 | 0 |
| 10 | 84 | 314 | 0 | 4 | 0 |
| 11 | 29 | 122 | 0 | 0 | 0 |
| 12 | 5 | 10 | 0 | 0 | 0 |
| 13 | 0 | 2 | 0 | 0 | 0 |
| 14 | 0 | 1 | 0 | 0 | 0 |
| **Total** | 1000 | 1000 | 1000 | 1000 | 1000 |

(The leftmost label column reads: Number of Line Rating Violations Per Sample)

In Fig. 11, we employed a scatter plot to display the objective total cost of the five proposed $N$-1 ROPF models against the objective total cost of the full $N$-1 OPF model. To enhance clarity without overwhelming the plot, we have confined the visualization to the initial 100 samples from our data set.

Notably, in Fig. 11, part (ii) and (iv), both the GNN model and HGNN model exhibit total costs lower than those of the full $N$-1 OPF model. This outcome comes from the breach of constraints in the $N$-1 scenarios. For the remaining models, the overall objective cost is around 0.25% higher in comparison to the full $N$-1 OPF model—a foreseeable result. This divergence can be attributed to our reduction of line constraints during the $N$-1 OPF formulation stage, which consequently relaxes the constraints governing generation ramping rates in the $N$-1 post-contingency scenarios. It is important to note that these ramping rates wield an impact on the total cost of each sample.

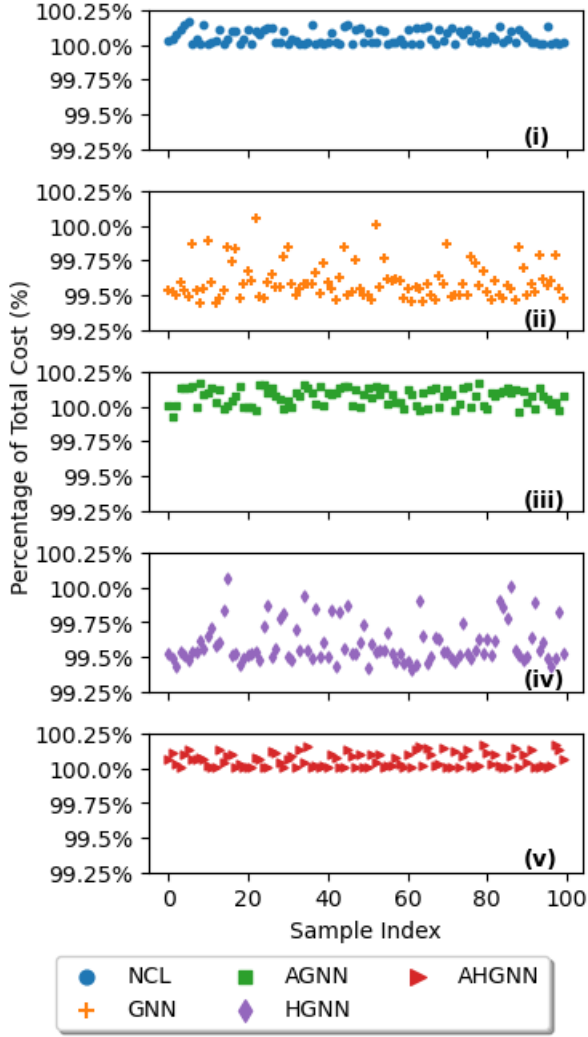

Fig. 11. Comparison of the objective total cost for the five proposed models.

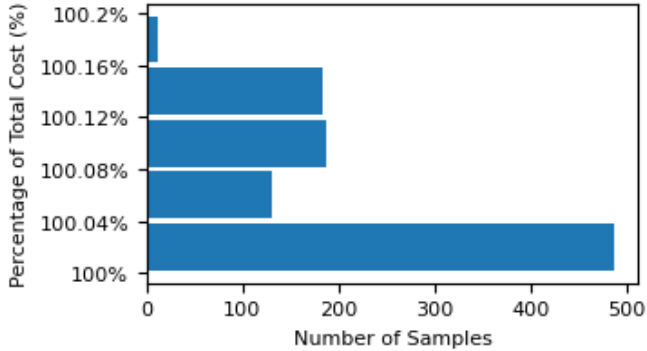

Fig. 12. Histogram of the total objective cost for AHGNN model.

Despite the variance in the total objective cost from the full *N*-1 OPF model, we truly believe that the modest increment has an inconsequential influence on the integrity of the *N*-1 ROPF solutions. As Table II outlines, the proposed AHGNN model manifests a maximum total cost of 100.193% compared to the full solution, while the median objective total cost stands at 100.043%. It signifies that the majority of *N*-1 ROPF solutions using AHGNN model contain an objective total cost with negligible increase compared to the full *N*-1 OPF solutions. Our assertion is validated in Fig. 12, where nearly

500 solutions register an objective total cost increase of less than 0.04%—a marginal and insignificant deviation.

| Model | NCL | GNN | AGNN | HGNN | AHGNN |
|--------|---------|--------|---------|--------|---------|
| **Mean** | 100.062 | 99.594 | 100.056 | 99.594 | 100.061 |
| **Median** | 100.043 | 99.551 | 100.046 | 99.551 | 100.043 |
| **Max** | 100.193 | 100.105 | 100.202 | 100.077 | 100.193 |
| **Min** | 100.002 | 99.365 | 99.919 | 99.365 | 100.002 |
| **Std.** | 0.052 | 0.135 | 0.060 | 0.135 | 0.051 |

### D. Solving Time Performance:

In the preceding sections, we conducted a comprehensive comparison of five benchmark models to evaluate their performance and efficacy in adhering to all constraints while staying close to the total objective cost. In the current section, we shift the focus to the computational performance analysis of our proposed model, AHGNN. Given that only the *N*-1 ROPF solutions obtained using the AHGNN model exhibit no line rating violations, these are the sole results that can be directly compared to the solutions of the full *N*-1 OPF.

To facilitate a comprehensive assessment, we solved the same data set of 1,000 samples using both the *N*-1 ROPF method and the full *N*-1 OPF method across five separate runs. Averaging the outcomes, we present the mean results in Fig. 13. The visualization showcases the total amount of time required to solve all 1000 samples. Specifically, the *N*-1 ROPF method demonstrated its capability by solving these 1000 samples in 7,098 seconds, reflecting a noteworthy 19% enhancement in speed as compared to the full *N*-1 OPF method.

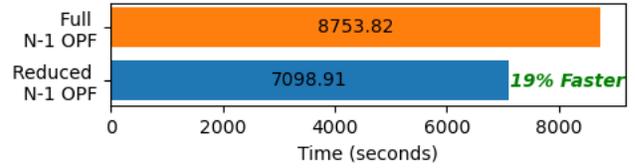

Fig. 13. Comparison of solving time between full *N*-1 OPF and *N*-1 ROPF.

While the reduction in the number of constraints was as high as 26%, going from around 84,000 down to 64,000, the actual decrease in computing time across all samples didn't exhibit a similar margin. The phenomenon can be attributed to the small size of the IEEE 73-bus system, containing just 108 branches. Consequently, the reduction in the number of constraints is not large enough, yielding a relatively proportional lesser decrease in computing time. However, it is worth noting that in larger systems with thousands of buses and a substantially higher number of branches, the gains in computing time reduction will likely rise exponentially.

This research paper specifically concentrates on the pragmatic application of the AHGNN model to expedite *N*-1 OPF computing times. The observed 19% reduction, though modest in percentage terms, is a significant achievement that aligns with our expectations and goals.

## VI. CONCLUSION

Our research has significantly reduced the number of constraints in the comprehensive *N*-1 ROPF problem. We achieved this by using the proposed innovative AHGNN model to formulate a streamlined *N*-1 ROPF algorithm that combines predictions from three distinct sources. This strategic integration of predictions resulted in a remarkable 19% decrease in solving time, while still preserving the quality of the solutions. We believe that by adopting a diverse strategic approach, we have designed an AHGNN model that can expedite solving times for *N*-1 ROPF problems.

For our current research, we focused on a relatively modest IEEE 73-bus system. However, we envisioned our research will be applied towards a larger-scale system, one that encompasses thousands of buses. We hypothesized that such a shift could result in substantial computational resource savings, a prospect that fuels our anticipation for the future trajectory of our research. For future projects, we plan to explore alternative topologies for GNN, with the ultimate goal of enhancing our predictive capabilities. Our methodology involves the utilization of diverse topologies for each bus, allowing us to intricately model each load and generator separately, thus achieving a more nuanced and detailed representation of the system dynamics. Looking forward, we plan to seamlessly integrate GNN into other advanced OPF problems, for instance, network reconfiguration OPF, thereby broadening the scope and applicability of GNN in power systems. This strategic expansion promises to unlock new insights and solutions in the realm of power system optimization, reinforcing the significance of ML technologies especially GNN.